\documentstyle[aps,prb,multicol]{revtex}

\draft

\input epsf
\input{epsf.sty}
\begin{document}

\hsize\textwidth\columnwidth\hsize
\csname@twocolumnfalse\endcsname

\widetext
\title{Strong pinning in high temperature 
superconducting films}

\author{C.J. van der Beek and M. Konczykowski}
\address{Laboratoire des Solides Irradi\'{e}s, CNRS-UMR 7642 and 
CEA/DSM/DRECAM, Ecole Polytechnique, 91128 Palaiseau Cedex, France}
\author{A. Abal'oshev, I. Abal'osheva,  P. Gierlowski, and S.J. 
Lewandowski}
\address{Instytut Fisyki Polskiej Akademii Nauk, Aleja Lotnik\'{o}w 32, 
02-668 Warszawa, Poland}
\author{M.V. Indenbom}
\address{Laboratoire des Solides Irradi\'{e}s, CNRS-UMR 7642 and 
CEA/DSM/DRECAM, Ecole Polytechnique, 91128 Palaiseau, France \\
and \\ Institute of Solid State Physics, Russian Academy of Sciences, 
142342 Chernogolovka, Russia}
\author{S. Barbanera}
\address{Istituto di Elettronica dello Stato Solido--Consiglio
Nazionale delle Ricerche, Via Cineto Romano 42-00156, Roma, Italy}

\date{\today}
\maketitle
\widetext
\begin{abstract}
Detailed measurements of the critical current density $j_{c}$ of 
YBa$_{2}$Cu$_{3}$O$_{7-\delta}$ films grown by pulsed laser deposition 
reveal the increase of $j_{c}$ as function of the film thickness. Both this thickness dependence and the 
field dependence of the critical current are consistently described using a generalization 
of the theory of strong pinning of Ovchinnikov and Ivlev [Phys. Rev. B {\bf 43}, 8024 (1991)]. 
From the model, we deduce values of the defect density ($10^{21} {\rm m^{-3}}$) and the 
elementary pinning force, which are in good agreement with the generally accepted values for 
Y$_{2}$O$_{3}$--inclusions. In the absence of clear evidence that the 
critical current is determined by linear defects or modulations of the 
film thickness, our model provides an alternative 
explanation for the rather universal field dependence of the critical current 
density found in YBa$_{2}$Cu$_{3}$O$_{7-\delta}$ films deposited by different methods.
\end{abstract}
\begin{multicols}{2}
\narrowtext

\section{Introduction}

High vortex pinning forces leading to high critical currents are an indispensable prerequisite for 
superconducting thin films if these are to be used in electronic and power 
applications. The ``volume'' pinning force exerted by material impurities
on flux vortices opposes their motion, which is at the origin of flux noise and dissipation. 
Strong vortex pinning decreases noise in electronic  
and superconducting quantum interference devices (SQUID's) and 
leads to high quality factors necessary for the correct operation 
of radio-frequency and microwave filters and cavities. Present-day films of the 
high temperature superconductor YBa$_{2}$Cu$_{3}$O$_{7-\delta}$  
combine high critical current density $j_{c}$ \cite{Roas91} with a high critical 
temperature $T_{c}$ and are thus ideal candidates for widespread 
application.  The optimization of the critical current 
through identification and tailoring of defect microstructures which 
lead to high pinning and reduced vortex creep has therefore attracted a lot of 
interest in the High Temperature Superconductor (HTC) community.

Several types of pinning defects have been suggested as being 
responsible for the large $j_{c}$'s in HTC films. Foremost are the oxygen 
vacancies in the strongly superconducting CuO$_{2}$ 
layers of these materials. Due to the small coherence length $\xi$ and the large condensation 
energy of the cuprates, such vacancies are potentially very effective pinning centers; their effect may
in principle explain the high critical currents encountered 
in YBa$_{2}$Cu$_{3}$O$_{7-\delta}$ films.\cite{Kes91,Griessen94}
Vortex pinning by these dense defects of typical diameter $D \sim 3 \AA \ll \xi$ is best 
described by the theory of weak collective flux pinning,\cite{Kes91,Larkin79,Blatter94} 
in which a net force on the vortex lattice exists only as a result of fluctuations in the defect 
density ($n_{d} \sim 10^{26} - 10^{27} {\rm m^{-3}} \gg 
\varepsilon \xi^{-3}$ is the effective defect density and $\varepsilon < 1$ is the anisotropy 
parameter of the superconductor, $\varepsilon \approx 0.14$ for 
YBa$_{2}$Cu$_{3}$O$_{7}$). 
The volume pinning force $F_{p}$ is obtained 
from $F_{p} = (WV_{c}^{-1})^{1/2}$, where the pinning strength $W = 
n_{d} \langle f_{p}^{2}\rangle$, $f_{p}$ is the elementary force 
exerted by one defect on a vortex line, and the averaging is carried 
out over a unit cell of the vortex lattice. The correlated volume $V_{c} = 
R_{c}^{2}L_{c}$, where the correlation lengths $L_{c}$ and $R_{c}$ 
are the distances, respectively parallel and perpendicular to the magnetic induction, at which the relative 
displacements due to pinning $\langle [u(R_{c},0) - u(0,0) ]^{2}\rangle^{1/2}$ and $\langle 
[u(0,L_{c}) - u(0,0) ]^{2}\rangle^{1/2}$ are equal to $\xi$.

The very high measured values $j_{c} \gtrsim 1 \times 10^{11} $ Am$^{-2}$ (see below) imply,  
however, that $R_{c}$ does not, under usual experimental conditions, 
exceed the vortex spacing $a_{0} = (2\Phi_{0}/\sqrt{3}B)^{1/2}$. 
\cite{Douwes} In other words, the small scale displacements of 
neighboring vortices are independent. An estimate of the longitudinal 
correlation length (or ``Larkin length'')  $L_{c}$ under the condition $R_{c} = a_{0}$, 

\begin{equation}
	L_{c} = \xi \left( \frac{ \sqrt{3} \varepsilon^{2} \varepsilon_{0} }{2 j_{c} 
	\Phi_{0}\xi } \right)^{1/2} \approx \varepsilon \xi \left( \frac{j_{0}}{j_{c}} \right)^{1/2},
\end{equation}

\noindent shows that $L_{c} \sim 10$ nm, much less than usual film 
thicknesses $d \gtrsim 100$ nm. Here $j_{0} = 
4\varepsilon_{0}/3\sqrt{3} \Phi_{0}\xi$ is 
the depairing current density, $\varepsilon_{0} = 
\Phi_{0}^{2}/4\pi\mu_{0}\lambda_{ab}^{2}$ is the typical vortex 
energy scale, $\Phi_{0} = h/2e $ is the flux quantum,  
$\lambda_{ab}$ is the penetration depth for supercurrents flowing 
in the CuO$_{2}$--layers, and $\mu_{0} = 4\pi \times 10^{-7}$. 
Contrary to the case of weakly pinning films in the two--dimensional 
limit, in which $j_{c} \propto 
(n_{d}/d)^{1/2}$,\cite{Kes81} the critical current density

\begin{equation}
	j_{c} = 
   	\frac{1}{\Phi_{0}} \left( \frac{n_{d} \langle  f_{p}^{2} \rangle \xi^{2} }{ L_{c}} \right)^{1/2} 
	\approx j_{0}      \left( 27 \frac{n_{d} \langle  f_{p}^{2} \rangle 
	\xi^{3} }{16 \varepsilon \varepsilon_{0}^{2}} \right)^{2/3}
	\label{eq:collective}
\end{equation}

\noindent is field-- and thickness--independent. Using the explicit form $\langle  f_{p}^{2} 
\rangle^{1/2} \approx  \frac{1}{4}\varepsilon_{0}(D_{v}/\xi)^{2}$ for oxygen 
vacancies, where $D_{v}$ is the oxygen ion radius, one has
	
\begin{equation}
j_{c} \approx j_{0} \left( \frac{ 27 n_{d}D_{v}^{4}}{256 \varepsilon \xi} \right)^{2/3}.
\end{equation}

\noindent Note, however, that underdoped twin-free YBa$_{2}$Cu$_{3}$O$_{7-\delta}$ single crystals, in 
which the increase of the sustainable current density due to an increasing density of oxygen 
vacancies in the CuO$_{2}$--layers as one underdopes the material is consistent with a 
concomitant decrease of $T_{c}$,\cite{Kokkaliaris2000} show an exponential temperature dependence of the 
current density that is not observed in thin films.

YBa$_{2}$Cu$_{3}$O$_{7-\delta}$ films typically form through two-dimensional 
nucleation and island growth,\cite{Hawley91,Gerber91} with lateral island sizes
that increase with growth 
temperature and film thickness.\cite{Raistrick93,Dam98} The significant thickness
modulations $\delta d$ related to the presence of the islands and 
the associated variations of the vortex line energy as function of the 
position, lead to pinning of vortices in the 
thinnest portions of the film. The maximum pinning force 
will be close to that needed to drive a single vortex line out of such 
a trough. If this mechanism is dominant, the 
experimentally measured critical current density is expected to follow

\begin{equation}
	j_{c}^{TV} 	\approx 
				\frac{2\pi\varepsilon_{0}}{\Phi_{0}D}\frac{\delta d}{d} 
			\approx  
				j_{0} \frac{3\sqrt{3} \pi \xi}{2 D}\frac{\delta d}{d} ,
	\hspace{1cm} 
	(B \ll \frac{\Phi_{0}}{D^{2}})
	\label{eq:thickness-variation}
\end{equation}

\noindent with $D$ the average island diameter.\cite{Kes91} In their
detailed study of the relation between film roughness and the 
current density, Jooss {\em et al.} suggested that this is indeed 
the case.\cite{Jooss96} However, the magnitude of the critical current 
density and its increase as function of the film thickness $d$ for 
films thinner than $2\lambda_{ab}$  could only be explained by these authors 
by invoking the lowered vortex line tension in  such thin
films, in which the effective penetration depth is equal to 
$2\lambda_{ab}/ d$, and by assuming an unidentified supplementary bulk 
pinning mechanism. Furthermore, the study\cite{Jooss96} restricted 
itself to fields $\mu_{0}H_{a} < 0.3$ T and $T = 5$ K. 

The discovery of growth spirals with a central screw dislocation in 
sputtered or metal-organic chemical vapor deposited (MO-CVD) YBa$_{2}$Cu$_{3}$O$_{7-\delta}$ films 
\cite{Hawley91,Gerber91} immediately lead to the suggestion that 
extended defects rather than microscopic point defects are the main 
pinning centers.\cite{Kes91} The growth spirals appear in laser-ablated films 
only when the substrate is heated above 850 $^{\circ}$C during  
deposition in an oxygen pressure exceeding 50 Pa, while below this temperature, 
the films are formed through island growth.\cite{Dam98} In the latter case, 
screw-- and edge dislocations are to be expected only in the troughs 
between islands, effectively halting easy vortex motion along these.
Recent experiments by Dam {\em et al.} \cite{Dam99} showed 
that a ``characteristic field'' (or vortex density) beyond which the critical 
current density of YBa$_{2}$Cu$_{3}$O$_{7-\delta}$ 
films decreases, is correlated with the surface density  of screw 
dislocations $n_{sd}$. Sequential etching showed 
the screw dislocations to extend throughout the film thickness, allowing for the 
possibility that the vortices are pinned along their entire length on the 
non-superconducting dislocation core of radius $c_{0} \ll \xi$.\cite{Dam99} 
The smallness of $c_{0}$ means that unlike amorphous columnar defects, 
to which the dislocation cores are often compared,\cite{Klaassen2002}
pinning is more likely to be due to the core-induced variation of 
the mean free path in the vicinity of the dislocation 
($\delta \kappa$--mechanism). Supposing that the pinning of an individual vortex 
line on a dislocation core leads to a critical current density 	
$j_{c}^{sd}(0)$, 
%
%
the critical current density at higher fields will be equal to 
$j_{c}^{sd}(0)$ times the fraction of pinned vortices $n_{t}$. 
The latter is determined by the probability that a given vortex 
encounters at least one dislocation core in an allowed area 
$U_{p}^{sd}/\varepsilon_{0} a_{0}^{2}$ determined by equating $U_{p}^{sd}$ 
to the loss of elastic energy due to the deformation of the vortex 
lattice,\cite{Wahl95,Hardy98,vdBeek2000} so that

\begin{eqnarray}
	j_{c}^{sd}  & = & j_{c}^{sd}(0)n_{t} \nonumber
	            \\
            	& = & j_{c}^{sd}(0) \left[ 1 - \left( 1 + \frac{U_{p}^{sd}}{\varepsilon_{0}} \right)
	\exp\left(-\frac{ \Phi_{0}n_{sd}U_{p}^{sd}}{\varepsilon_{0} B}\right) \right].
	\nonumber
	\\
	\label{eq:SD-high-field}
\end{eqnarray}

\noindent A similar formula was used by Dam {\em et al.} to deduce 
the value of the ``characteristic field'' $B^{*} \equiv \Phi_{0} n_{d}
U_{p}^{sd}/\varepsilon_{0}$. 
Due to the non-homogeneous current distribution in 
superconductors, depinning from the linear defects would be initiated by 
the nucleation at the film surface of vortex kinks joining two 
dislocation cores.\cite{Prozorov98} 
As a result, the experimentally measured low--field ``critical'' current 
$j^{sd}(0)$ should be rather smaller than the critical current density 
in the absence of flux creep and decay rapidly with time. 
The time decay will be even more rapid when the applied magnetic 
field and the vortices are not aligned along the dislocation cores. Hence, in analogy to the 
case of heavy-ion irradiation--induced amorphous tracks, pinning by extended linear screw 
dislocation cores should lead to a sharp cusp-like maximum in the 
field orientation--angle dependence of the experimentally measured 
critical current.\cite{Nelson92,Nelson93} 

It should be remarked that both in the case of pinning by thickness 
variations and pinning by screw dislocations, the low--field critical
current flows only at the film surface. Hence, the total 
screening current as well as the characteristic field of 
flux penetration $H_{c}$ in a magnetic experiment (see below) is independent of the film 
thickness $d$, and the apparent current density $j$ decreases inversely 
proportional to $d$.

In what follows, we present a detailed study of the critical current density of 
YBa$_{2}$Cu$_{3}$O$_{7-\delta}$ films grown by pulsed laser 
deposition as function of film thickness, field magnitude and orientation, 
and temperature. It turns out that 
the field orientation-- and thickness dependence of $j_{c}$ do not provide any 
evidence for pinning by screw dislocations or other correlated 
disorder in our films. Yet, the temperature and field dependence of 
$j_{c}$ is entirely comparable to that measured in films which contain 
these defects.\cite{Gerber91,Dam99} 

To explain our results, we propose that the relevant pinning centers in our 
films are sparse insulating or normal metallic second-phase inclusions. 
These are known to exist in sputter-deposited YBa$_{2}$Cu$_{3}$O$_{7-\delta}$ 
films as platelet-like Y$_{2}$O$_{3}$ inclusions of size $15 \times  15 \times 10$ nm$^{3}$, 
with typical densities $n_{i} \sim 10^{22} - 10^{23} {\rm m^{-3}}$.\cite{Selinder92,Catana} Other 
authors have reported the presence of Y$_{2}$O$_{3}$ inclusions in 
laser--ablated films as well, with typical densities $n_{i} \approx 10^{22}$ m$^{-3}$ 
for inclusions of diameter $D_{i} \sim 3$ --- 5 nm, and $n_{i} \approx 10^{21}$ 
m$^{-3}$ for $D_{i} \sim$ 10  -- 20 nm.\cite{Kastner95,Verbist96}  
Pinning by such large defects turns out to be conveniently described using an extension of 
the theory of strong pinning of Ovchinnikov and Ivlev 
\cite{Ovchinnikov91}, which we shall develop below  
(\ref{section:theory}). Our model is, to our knowledge, the only that
consistently describes both the field- and thickness dependence of the 
critical current density in YBa$_{2}$Cu$_{3}$O$_{7-\delta}$ films.

\section{Strong pinning by sparse large point pins}
\label{section:theory}

\subsection{General formulation}

Strong pinning by large point defects has been 
described by simple substitution of the elementary pinning force of a large 
void $f_{p,max}^{i}$ into the collective 
pinning expression (\ref{eq:collective}).\cite{Douwes} This 
procedure makes the implicit assumption that there are many 
inclusions in a region of volume $a_{0}^{2}L_{c}$. A comparison of 
the longitudinal correlation length, $L_{c} \sim 10$ nm, which follows from a 
collective-pinning analysis of experimental critical current 
density--data, with the expected mean distance between large
defects, $d_{i} \sim 30$ nm, shows that this approach is inappropriate.

In order to obtain the critical current density for sparse pins, one 
should not start from the statistical average of the pinning forces 
of the different defects, but, rather, evaluate the probability that a vortex 
line will be pinned at all. A nearby defect will be able to trap a 
vortex line if the gain in pinning energy $U_{p}^{i} \approx 
f_{p,max}^{i}\xi$ is sufficient to outweigh the elastic energy loss 
due to the vortex lattice deformation ${\bf u}({\bf r})$ caused by displacing the 
vortex line onto the defect. The maximum allowed lateral displacement of 
the vortex $u_{0}$ determines the ``trapping area'' $\sim u_{0}^{2}$ 
within which a large defect is an effective pinning site. The bulk 
(volume) pinning force $F_{p} = (a_{0}^{2}d)^{-1}\sum_{i}^{N} f_{p,max}^{i}$ 
is given by the direct sum of the elementary forces of the individual defects 
that can effectively pin a single vortex, \em i.e. \rm those that lie 
within $u_{0}$ of the vortex lattice position, normalised by the volume 
$a_{0}^{2}d$ available to the vortex. Any vortex line will be pinned 
on average by $N = d/ \overline{\cal L}$ defects, where 
$\overline{\cal L}$ is the average distance between such ``effective'' 
pinning centers, so that

\begin{equation}
	F_{p} = \frac{f_{p,max}^{i}}{a_{0}^{2}d} \frac{d}{\overline{\cal L}} = \frac{f_{p,max}^{i}}{a_{0}^{2} \overline{\cal L} }.
	\end{equation}
	
\noindent 
The probability to encounter a second effective 
defect at distance $\cal L$ from a first defect located at $z=0$ is given by the product of the 
probability to encounter none in the interval $0<z<{\cal L}$ and the probability to find 
at least one at $z={\cal L}$.  In the case of layered superconductors  
considered in Ref.~\onlinecite{Ovchinnikov91}, where the only effective defects are 
those situated in the CuO$_{2}$ layers, this yields 

\begin{eqnarray}
	\overline{\cal L} & = & s + s \left[ 1 - \exp \left( - n_{\Box}u_{0}^{2} \right) \right]
	                        \sum_{k=1}^{\infty} k \exp \left( -k  n_{\Box} u_{0}^{2} \right)
			\nonumber	\\
				 & = & \frac{s}{  1 - \exp \left( -n_{\Box} u_{0}^{2} \right) }
	\label{eq:strong-layered}
\end{eqnarray}
	
\noindent and $j_{c} = F_{p}/ B = ( f_{p,max}^{i}/\Phi_{0}s ) [ 1 - \exp ( - n_{\Box} 
u_{0}^{2} ) ]$. Here $n_{\Box}$ is the areal density of defects 
in the CuO$_{2}$ layers and $s$ is the distance between layers. This 
result has the same ``two--dimensional'' form as the result ( \ref{eq:SD-high-field} )
for  linear defects extending throughout the thickness of the film. For 
continuous\cite{Blatter94}  superconductors

\begin{equation}
	\overline{\cal L}  =  \int_{0}^{\infty} {\cal L} \exp \left( - n_{i} 
	u_{0}^{2} {\cal L} \right) d{\cal L}
	                    =  \frac{1}{n_{i}u_{0}^{2}}  
	\label{eq:strong-continuous}
	\end{equation}
	
\noindent which gives \cite{Ovchinnikov91}

\begin{equation}
	j_{c} = \frac{f_{p,max}^{i}}{\Phi_{0}} n_{i} u_{0}^{2}.
	\label{eq:strong}
\end{equation}
 
\noindent In the case of real samples (of finite thickness in the 
field direction), and especially for thin films, the derivation (\ref{eq:strong-continuous}) 
becomes inappropriate in the limit $n_{i}u_{0}^{2}d  \ll 1$, in which not every vortex line can find 
a defect. For such small defect densities or sample thicknesses, one should evaluate 
the probability ${\cal P} = 1-\exp( -n_{i}u_{0}^{2}d )$ that a vortex line 
encounters at least one defect in the volume $u_{0}d$. However, the same condition 
$n_{i}u_{0}^{2}d  \ll 1$ implies  ${\cal P } \approx n_{i}u_{0}^{2}d$; Hence, the critical current 
density $j_{c}  =  (f_{p,max}^{i}/\Phi_{0}d) {\cal P} $  has the same form 
as Eq.~(\ref{eq:strong}).
We shall, therefore, assume that Eq.~(\ref{eq:strong}), valid for both 
$n_{i}u_{0}^{2}d  \ll 1$ and $n_{i}u_{0}^{2}d  \gtrsim 1$, is applicable irrespective of the defect density and the film 
thickness.\cite{NB}

\subsection{The trapping area}

\subsubsection{Pin Breaking}
\label{sec:pinbreaking}

The trapping area and $f_{p,max}^{i}$ can be obtained using two different 
approaches, corresponding to the different mechanisms by which a vortex 
can be liberated from a defect: pin-breaking and plastic 
depinning.\cite{Schonenberger96}
The most obvious possibility is ``pin-breaking'',
in which the applied force must exceed the attractive force $f_{p,max}^{i} $
exerted by a defect on a vortex line. Plastic depinning only occurs 
if the pins can be considered as ``infinitely strong'', \em i.e. \rm 
$f_{p,max}^{i} \gg \varepsilon_{0}$, which is not the case in what 
follows. Exploiting the similarity of 
pinning by extended pointlike defects 
with pinning by amorphous columnar defects,\cite{Nelson93}
the elementary pinning force of an inclusion of extent (perpendicular to the field direction) 
$D_{i}$ can be estimated as the product of the fraction of the vortex core 
volume occupied by the defect and the condensation energy 
$B_{c}^{2}/2\mu_{0} = \frac{1}{4}\varepsilon_{0}\xi^{-2}$. 
%
Following Ref.~\onlinecite{Blatter94}, the resulting pinning 
force can be approximated by the interpolation

\begin{eqnarray}
	f_{p,max}^{i} & \approx &  \varepsilon_{0} \left( \frac{D_{i}^{z}}{4\xi} \right) 
	\ln \left( 1 + \frac{D_{i}^{2}}{2\xi^{2}} \right)  \nonumber \\
	& \equiv &  \varepsilon_{0} \left( \frac{D_{i}^{z}}{4\xi} \right)
	{\cal F}(T) ,
	\label{eq:fp}
\end{eqnarray}

\noindent where $D_{i}^{z}$ is the extent of the defect 
along the field direction. Note that the above expression corresponds 
to ``$\delta T_{c}$--pinning'';\cite{Blatter94} pinning by the variation
of the mean-free path in the vicinity of the defect is relatively 
unimportant, because the quasiparticle scattering probability in the 
layer of thickness $\xi_{0}$ surrounding the defect is 
negligible as compared to the total scattering cross-section of the 
inclusion ($\xi_{0}$ is the BCS coherence length).

The maximum allowed vortex displacement $u_{0}$ is 
obtained by balancing the elastic energy loss with the pinning energy 
gain $U_{p} \sim f_{p,max}^{i}\xi$. Taking explicitly into account the range 
of the pinning potential , Ovchinnikov and Ivlev obtained 
\cite{Ovchinnikov91}

\begin{equation}
	u_{0}^{2}  =  \left(\frac{128}{27}\right)^{1/4} \left( \frac{U_{p}}{\varepsilon\varepsilon_{0}}\right)^{5/4} 
	 	\frac{a_{0}^{5/4}}{\xi^{1/2}}  .  
				\label{eq:area-Ovchinnikov}
\end{equation}

\noindent Here we have approximated the vortex lattice shear modulus
$c_{66} \approx \frac{1}{4}\varepsilon_{0}a_{0}^{-2}$ and the 
nonlocal tilt modulus $c_{44} \approx \varepsilon^{2}\varepsilon_{0}a_{0}^{-2}$.
Note that for fields in excess of $B_{a} \equiv 0.41 \Phi_{0} \xi^{4/3} ( 
\varepsilon\varepsilon_{0}/U_{p})^{10/3}$
the maximum allowed vortex excursion becomes comparable to $a_{0}$, and $u_{0}^{2} 
\approx a_{0}^{2}$.

The critical current density follows from combining 
Eqs.~(\ref{eq:strong}), (\ref{eq:fp}), and (\ref{eq:area-Ovchinnikov}), 

\begin{eqnarray}
	j_{c} & \approx & 0.0866 n_{i} j_{0} 
                       \frac{\left[  D_{i}^{z} {\cal F}(T) \right]^{9/4}}
						            { \varepsilon^{5/4} \xi^{1/2} } 
	                    \left( \frac{\Phi_{0}}{B} \right)^{5/8} 
		\hspace{0.15cm} ( B \ll B_{a} ) 
		\label{eq:jc-strong}  \\
	j_{c} & \approx & 0.375 n_{i} j_{0} D_{i}^{z}{\cal F}(T)  \frac{\Phi_{0}}{B}    
		\hspace{2.1cm} ( B \gg B_{a} ).
		\label{eq:jc-strong-high-field}
\end{eqnarray}

The above expression can be compared to the result of a simpler 
estimate, which follows from an 
approach initially suggested by Vinokur 
{\em et al.} to describe pinning by dense point pins and vortex wandering in 
layered superconductors.\cite{Vinokur98} The energy of elastic deformation of
a representative vortex segment of length $L$ in its lattice 
cell is $U_{el} = c_{66}u^{2}L + c_{44} (u^{2}/L)a_{0}^{2}$. 
Minimization with respect to $L$ yields the optimum length $L_{0} =  
a_{0}(c_{44}/c_{66})^{1/2} \approx 2\varepsilon a_{0}$ over which a vortex segment can 
fluctuate independently 
from its neighbors. Note that $L_{0}$ is equal to $L_{c}$ on condition that
$R_{c} = a_{0}$ exactly. The distance $u_{0} = u(L_{0})$, to which a vortex may wander, is 
obtained by equating the pinning energy gain $U_{p}$ to $U_{el}(u_{0},L_{0})$; this gives 

\begin{equation}
	u_{0}^{2} =  \frac{U_{p}}{\left(c_{66}c_{44}\right)^{1/2} a_{0}} \approx 
	             \frac{ U_{p} }{ \varepsilon\varepsilon_{0} }  a_{0}.
	\label{eq:area}
\end{equation}

\noindent The critical current density 

\begin{eqnarray}
    j_{c} & = & 0.0875 n_{i} j_{0} 
                 \frac{ \left[ D_{i}^{z}{\cal  F}(T) \right]^{2}} {\varepsilon }
            	\left( \frac{\Phi_{0}}{B} \right)^{1/2} 
			\hspace{0.5cm} ( B \ll \tilde{B}_{a} ) 	
			\label{eq:jc-cage} \\
	     & = & 0.375 n_{i} j_{0}  D_{i}^{z}{\cal  F}(T) \frac{\Phi_{0}}{ B }
		    \hspace{2.2cm} ( B > \tilde{B}_{a}).
            \label{eq:cage-high-field}
\end{eqnarray}

\noindent with $\tilde{B}_{a} =\Phi_{0} (\varepsilon \varepsilon_{0} / U_{p} )^{2}$.
 
For convenience, we shall leave Eqs.~(\ref{eq:jc-strong}), 
(\ref{eq:jc-strong-high-field}) and 
(\ref{eq:jc-cage}), (\ref{eq:cage-high-field}) in their general form, since 
this permits one to take the possible  effects of thermal smearing of the pinning 
potential \cite{Nelson93,Feigel'man90} into account by simply 
replacing ${\cal F}(T)$ by the appropriate temperature dependence. 


\subsubsection{Low fields - single vortex limit}
\label{section:SV}

A glance at Eqs.~(\ref{eq:area-Ovchinnikov})--(\ref{eq:cage-high-field}) shows that at 
low fields the trapping area and  hence the critical current have an unphysical divergence.
The correct low--field limit of the trapping area is obtained by 
starting from the line tension of a single vortex line 
$\varepsilon_{1} \approx \varepsilon^{2}\varepsilon_{0}$. Balancing the 
energy of elastic deformation of a single line with the pinning 
energy, $\varepsilon_{1}u^{2}/{\cal L} = U_{p}$, one has $u^{2} = 
(U_{p}/\varepsilon_{1}){\cal L}$ and 

\begin{eqnarray}
	\overline{\cal L} & = &  \frac{ \int_{0}^{\infty} {\cal L} \exp \left( n_{i} U_{p} {\cal L}^{2}/ \varepsilon_{1} \right) d{\cal L} }
	                              { \int_{0}^{\infty}   \exp \left( n_{i} U_{p} {\cal L}^{2}/ \varepsilon_{1} \right) d{\cal L} }
						\nonumber  \\
				&	 = & \frac{1}{\sqrt{\pi}} \left( \frac{\varepsilon_{1}}{n_{i}U_{p}} \right)^{1/2}.
				\label{eq:L-SV}
\end{eqnarray}

\noindent The trapping area becomes $u_{0}^{2} = ( U_{p}/ \pi n_{i} \varepsilon_{1})^{1/2}$ and 

\begin{eqnarray}
	j_{c} &  =      &           n_{i}^{1/2} 
	\frac{f_{p,max}^{i}}{\Phi_{0}} \left( \frac{\pi U_{p}}{\varepsilon^{2}\varepsilon_{0}} \right)^{1/2} 
	\nonumber \\
	      & \approx & 0.28 n_{i}^{1/2} j_{0}    \frac{\left[ D_{i}^{z}{\cal  F}(T)\right]^{3/2} }{2\varepsilon}  
		 \label{eq:jSV}
\end{eqnarray}

\noindent The single vortex limit is realized for fields such that
$( U_{p}/\pi n_{i}\varepsilon_{1})^{1/2} \lesssim
(U_{p}/\varepsilon\varepsilon_{0}) a_{0}$, \em i.e. \rm $B \lesssim B^{*} 
\equiv  \pi \Phi_{0}  n_{i}  ( U_{p}/ \varepsilon_{0} )$.

\subsubsection{Very thin films}

A second limit, in which the results of Section~\ref{sec:pinbreaking} do not hold is that of very 
thin films of thickness $d <  L_{0}$. Note that this condition is also 
violated at very low fields $B \ll \Phi_{0} \varepsilon / d^{2}, \Phi_{0} /\lambda_{ab}^{2}$. 
In YBa$_{2}$Cu$_{3}$O$_{7-\delta}$ we have $L_{0}$($B = 10$  mT) $\approx 150$ nm, 
which is comparable to typical film thicknesses. Under these conditions, the probability for a 
vortex to be trapped by an inclusion is determined by its ability to bend 
sufficiently within the film thickness, {\rm i.e.} the tilt contribution 
dominates the elastic energy. We thus need to repeat our 
considerations for $L_{0} = d$, which will minimize the elastic 
energy in this case. The total energy of a vortex in the cage
of dimensions $a_{0}^{2}d$ is given by

\begin{equation}
	c_{66} u^{2}d + c_{44}  \frac{u^{2}}{d}  a_{0}^{2} - U_{p} 
	\approx  \varepsilon_{1} \frac{u^{2}}{d} - U_{p}.
	\label{eq:elastic-film}
\end{equation}

\noindent Equating this to zero yields $u_{0}^{2} = d (U_{p}/\varepsilon_{1})$, 
and the critical current density 

\begin{equation}
	j_{c} = n_{i} \frac{f_{p,max}^{i}}{\Phi_{0}}  \frac{U_{p}d}{\varepsilon^{2}\varepsilon_{0}}.
	\label{eq:jthinthin}
\end{equation}
	
\noindent Thus, for very thin films, the critical current density 
should be field-independent and increase linearly with film thickness.

Under the condition that $B < B^{*}$, the crossover from the ``thick thin film'' 
single vortex limit [section (\ref{section:SV})] and the ``thin thin film'' 
limit occurs when the thickness is reduced below the crossover 
value 

\begin{equation}
	d^{*} = \left( \frac{\varepsilon^{2}\varepsilon_{0}}{\pi n_{i} U_{p}} \right)^{1/2}
\label{eq:d*}
\end{equation}

\noindent determined from the equation of the trapping areas 
$u_{0}^{2}$ in either limit.

%
%
%
%


\section{Experimental details}

Thin films of YBa$_{2}$Cu$_{3}$O$_{7-\delta}$ were deposited on 
1.2$\times$1.5 mm$^{2}$ LaAlO$_{3}$ substrates by KrF
laser ablation from a stoichiometric target.   The 
deposition was carried out onto  substrates heated to 785$^{\circ}$ 
C at 280 mTorr (36.7 Pa) oxygen pressure, in the so-called ``off-axis'' geometry,
\em i.e. \rm the substrate surface was oriented parallel to the plasma
plume axis. From previous experiments, we know this geometry to 
produce films featuring homogeneous flux penetration (on a $\mu$m 
scale). We have prepared different series of films of seven different 
thicknesses, from 100 to 500 nm, using exactly the same 
deposition procedure. After deposition, the 
film edge was etched away to produce 0.9$\times$0.9$\times$mm$^{2}$ 
YBa$_{2}$Cu$_{3}$O$_{7-\delta}$ squares used for further investigations.

The obtained films were characterized using X-ray diffraction with 
Cr$_{K \alpha}$ radiation and Atomic Force Microscopy (AFM). The 
diffraction experiments revealed

\begin{figure}
\centerline{\epsfxsize 8.5cm \epsfbox{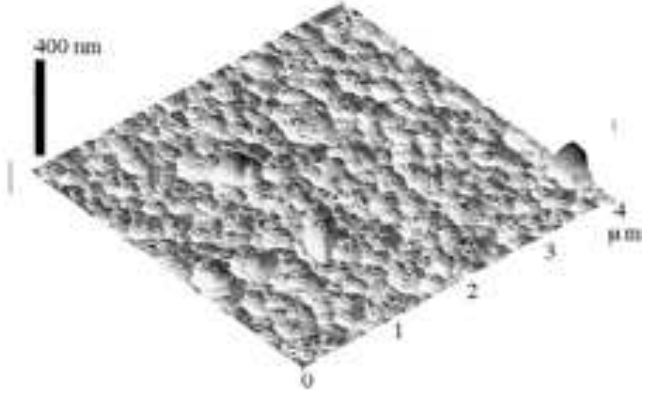}}
\vspace{1mm}
\caption{\label{fig:AFM}
 Tapping mode  AFM image showing the surface morphology of a
laser-ablated YBa$_{2}$Cu$_{3}$O$_{7-\delta}$ thin film of 300 nm thickness.}
\end{figure}

\noindent the existence (5 atomic \% ) of a second 
(cubic) phase. In AFM experiments we observed a surface morphology characteristic 
of island growth, with occasional outgrowths (see Fig.\ref{fig:AFM}).
The growth islands had a typical lateral size of about 300 nm.
This is very similar to the results found by other authors under the same 
deposition conditions.\cite{Dam98,Dam99} We do not find the spiral 
growth mechanism to be relevant to our samples -- a comparison with the 
deposition conditions of Ref.~\onlinecite{Dam98} indicates that growth 
spirals are not to be expected.

All films were characterized using the magneto-optical technique for flux 
visualization.\cite{Dorosinskii92} A ferrimagnetic garnet film with 
in-plane anisotropy is placed directly on top of the superconducting 
film. The garnet is observed using a polarized light microscope with 
nearly crossed polarizers. In this configuration, the reflected light 
intensity is proportional to the local magnetic induction 
perpendicular to the garnet, allowing the direct observation 
of flux penetration into the YBa$_{2}$Cu$_{3}$O$_{7-\delta}$ film.
The ``critical'' screening current density $j$ can 
be obtained directly from the distance $x_{f}$ between the flux front 
and the film edge at a given applied magnetic field $H_{a}$ and 
temperature, $j = \pi H_{c}/d =  \pi H_{a}/ [ d {\rm arccosh}( w / 
w-x_{f} )]$, where $w$ is half the film width and $H_{c} = j d / \pi$ the 
``characteristic field'' for flux penetration.\cite{Brandt93} 
Alternatively, one can fit the  position of the flux front as function 
of the applied field $H_{a}$, 

\begin{equation}
	x_{f} =  w \left[ 1 - \frac{1}{ \cosh \left(  \pi H_{a}/j d   
	\right)} \right] ,
	\label{eq:fluxfront}
	\end{equation}	
	
\noindent to obtain the low-field limit of the current density $j$. \cite{Jooss96}

The magneto-optical imaging of the films was used not only to obtain current densities 
at fields smaller than 300 G, but also to select the 
most homogeneous films for magnetometry experiments using the Local Hall 
Probe  Magnetometer.\cite{Konczykowski91II}
A commercial miniature GaAs

\begin{figure}
\centerline{\epsfxsize 8.5cm \epsfbox{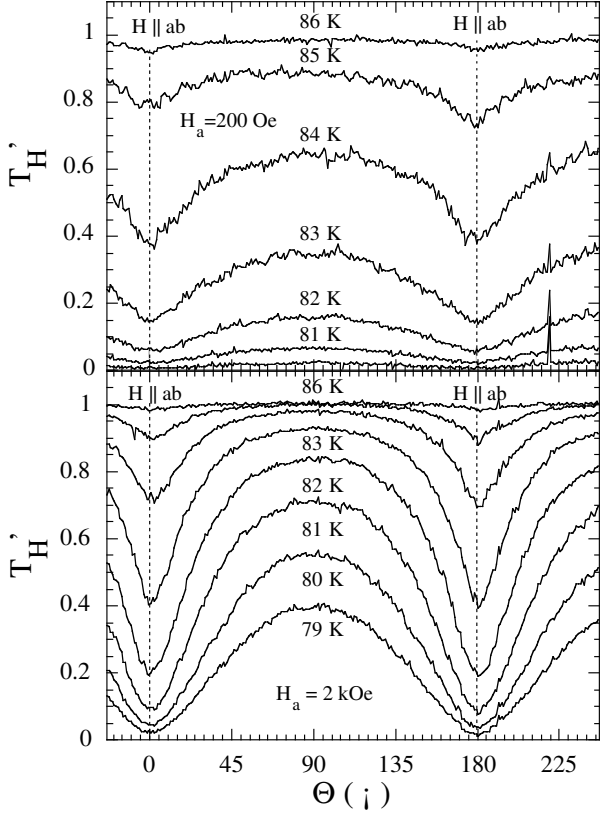}}
\vspace{1mm}
\caption{\label{fig:vierge-angulaire}
Field-orientation dependence of the in--phase fundamental transmittivity 
for a laser-ablated   YBa$_{2}$Cu$_{3}$O$_{7-\delta}$ thin film 
250 nm thick, for fields of 200 Oe (a) and 2 kOe (b). The 
temperature is decreased from 86 K down to 79  K in 1 K steps. The 
ac field had an amplitude of 1.5 Oe and a frequency of 7.75 Hz.}

\end{figure}

\noindent Hall probe, placed in the center of the 
film surface, is used to measure the local induction as function of 
the applied magnetic field. The result is converted into the so-called 
``self--field'' of the sample, $H_{s} = B - H_{a}$, created by the 
circulating screening current. The latter was determined from the 
width of the  hysteresis loop of $H_{s}$ at constant $B$ and calibrated 
using the values obtained from the magneto-optic experiments. In this 
way, the temperature-- and field dependences of the screening current 
were measured in fields of up to 2 T and temperatures between 30 and 85 
K. In order to obtain results in the temperature range close to 
$T_{c}$, measurements in AC mode were carried out by applying a 1.5 Oe AC field of 
frequency 7.75 Hz. The AC 
Hall voltage $V(f)$, measured using a dual-reference lock-in amplifier, was converted
to the in-phase fundamental transmittivity of the sample, 
$T_{H}^{\prime} = [V(f, T) - V(f, T \ll T_{c})] /  [V(f, T \gg T_{c}) - V(f, T \ll 
T_{c})]$.\cite{Gilchrist93} Using this technique, the critical temperature of all 
films was found to be similar, $T_{c}$ ranging between 88.8 and 91.7 K. 

 \begin{figure}
\centerline{\epsfxsize 8.5cm \epsfbox{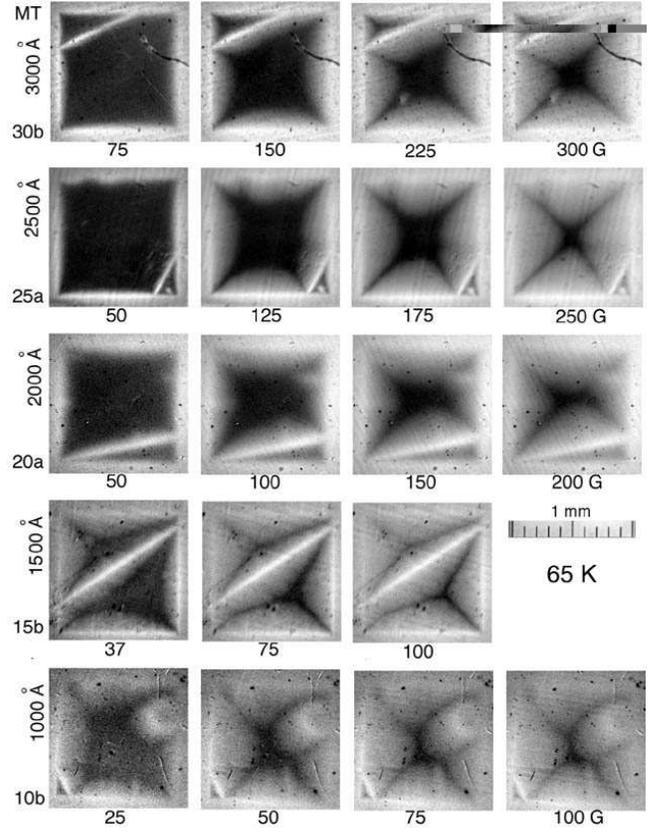}}
\vspace{4mm}
\caption{\label{fig:compare65K}
Magneto-optical images of flux penetration into laser-ablated 
YBa$_{2}$Cu$_{3}$O$_{7-\delta}$ films of 
thickness $d = 300$ nm, 250 nm, 200 nm, 150 nm and 100 nm 
respectively, at 65 K. Bright areas correspond to a 
high magnetic induction perpendicular to the film, while dark regions 
are areas of  low flux density. Horizontal sequences 
correspond to constant film thickness and varying applied field $H_{a}$ 
(the values are indicated below each frame), while the vertical sequences are 
deliberately chosen such that $H_{a} / d$ is approximately  
constant ---  a procedure, which reveals the weaker pinning in the 
thinner films. This is most clearly seen for the lower fields. Weak 
links in the films are revealed as the bright lines of preferential 
flux penetration.}
\end{figure}
 
\section{Results}

Transmittivity experiments as function of DC field orientation 
performed at different temperatures and field  magnitudes do not  
reveal any cusp in the angular dependence of the transmittivity when 
$H_{a}$ is oriented along the film normal (Fig.~\ref{fig:vierge-angulaire}).
This indicates that it is unlikely that directional pinning by correlated disorder 
determines the critical current density. If such pinning is 
due to the dislocations cores, it is masked by another mechanism of strong 
pinning, which remains to be identified.

In order to evaluate the thickness dependence of the critical current 
density without ambiguities in the conversion of magnetic moment to 
current density, we turn

\begin{figure}
\centerline{\epsfxsize 8.5cm \epsfbox{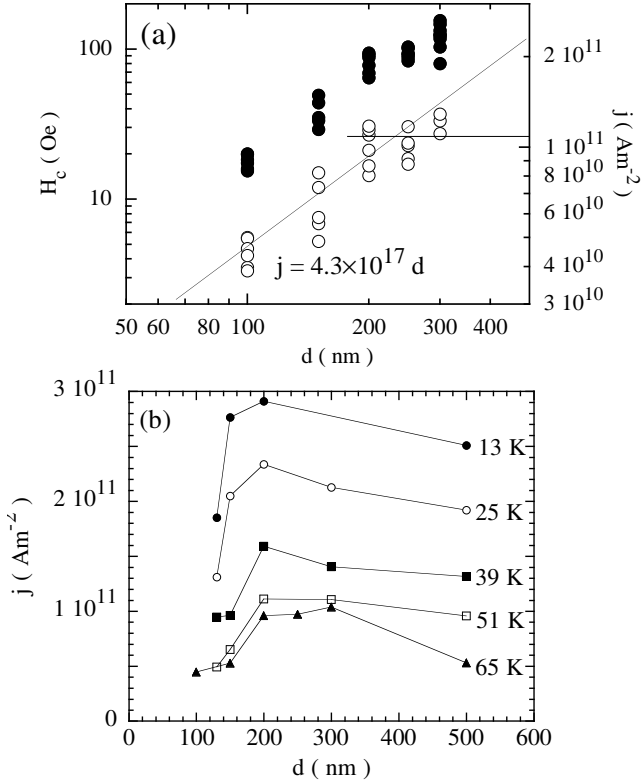}}
\vspace{1mm}
\caption{\label{fig:jc-vs-thickness}
(a) Dependence on film thickness $d$  of the characteristic field 
$H_{c}$ ( $\bullet$ ) and the screening current density $j = 
H_{c}/d$ ( $\circ$ ) of laser-ablated YBa$_{2}$Cu$_{3}$O$_{7-\delta}$ 
films in the limit of small flux densities $B$, at $T = 65$ K. Data 
points were obtained from inserting the position $x_{f}$ of the flux front at 
various applied fields in $j = \pi H_{c}/d =  \pi H_{a}/ [ d {\rm 
arccosh}( w / w-x_{f}/w )] $. The drawn lines indicate the linear increase $j \sim 4.3\times 10^{17} d$
for  $d \protect\lesssim 200$ nm, and the plateau reached for larger thicknesses. This
behavior is in agreement with the predictions (\protect\ref{eq:jSV}) and
(\protect\ref{eq:jthinthin}). (b) Film thickness dependence of the 
screening current density for various temperatures. Data points were 
obtained by fitting the position of the flux front as function of 
applied field to Eq.~(\protect\ref{eq:fluxfront}) }
\end{figure}

\noindent  to magneto-optical imaging of the flux 
penetration. Fig.~\ref{fig:compare65K} shows the 
flux penetration into five films of thicknesses $100 < d < 300 $ 
nm, at $T = 65$ K.  Nearly all films of this batch show at least one weak 
link, possibly due to the presence of a grain boundary in the 
substrate. These are easily rendered visible due to the preferential flux 
penetration along them. Nevertheless, the remaining areas feature homogeneous
flux penetration and are sufficiently large to obtain reliable results for the flux 
front position and the screening current density. The image frames 
are chosen so that the vertical sequences have $H_{a}/d \approx$ 
constant,  a procedure that should yield similar flux penetration if
the screening current would be the same for all films. It is clearly 
seen, however, that for constant $H_{a}/d$ flux penetration is easier 
in the thinner films. The effect is quantified by measuring the position 
of the flux-front relative to the film edge and  converting

\begin{figure}
	\centerline{\epsfxsize 8.5cm \epsfbox{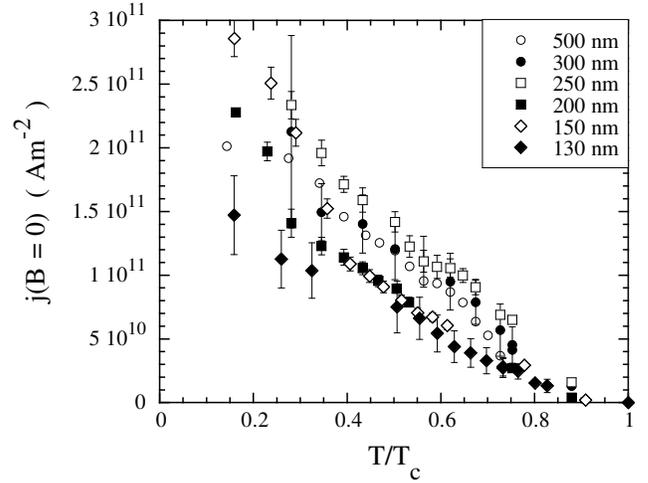}}
	\caption{Compilation\label{fig:jc(0)-compilation} of $j_{c}(B = 0)$--values for 20 films of 
	thickness $130 < d < 500$ nm obtained from different growth batches. The data 
	points show the average current density measured on films of the same 
	thickness, while error	bars indicate the rms deviation in $j$. }
	\end{figure}

\noindent this to the characteristic field $H_{c}$ and screening current density
$j$ using Eq.~(\ref{eq:fluxfront}). The result of this procedure is
plotted in Fig.~\ref{fig:jc-vs-thickness}(a). We find that $j$ manifestly increases 
before reaching a plateau for $d \gtrsim 200$ nm. This increase and 
the subsequent plateau is in contradiction both with the constant $j(d)$ predicted 
for weak collective pinning (Eq.~(\ref{eq:collective})), and with the constant 
$H_{c}$ and $j \propto d^{-1}$ expected for pinning by linear defects 
or by thickness variations (Eq.~(\ref{eq:thickness-variation})).
  However, it is in agreement with  
Eqs.~(\ref{eq:jSV}) and (\ref{eq:jthinthin})  derived for sparse inclusions or large 
point pins, which predict that the sustainable current should increase with 
thickness before reaching a plateau for $d \gtrsim d^{*}$. 
The experimental data show a slope $4.3 \times 10^{17}$, which can be 
compared to the theoretical value  
$n_{i}(f_{p,max}^{i}/\Phi_{0})( U_{p}/\varepsilon^{2}\varepsilon_{0})$ 
predicted by Eq.~(\ref{eq:jthinthin}). This yields,
at 65 K,  a defect density $n_{i} = 1\times 10^{21}$ m$^{-3}$, if we 
use   Eq.~(\ref{eq:fp}) for the pinning force,  $D_{i} = 
15$ nm and $D_{i}^{z} = 10$ nm from TEM experiments, 
\protect\cite{Selinder92} $\varepsilon_{0} = 1.7\times 10^{-11} 
(1-t^{4})$ Jm$^{-1}$, $\xi = 2 [(1+t^{2})/(1-t^{2})]^{1/2}$ nm,
$t \equiv T/T_{c}$ and $T_{c} = 91.7$ K. The obtained defect density 
corresponds to a crossover thickness $d^{*} = 60$ nm at 65 K, comparable 
to the experimental data.

In order to check the plausibility of the above results, we have 
prepared two other series of YBa$_{2}$Cu$_{3}$O$_{7-\delta}$ thin 
films using the same nominal deposition conditions. All 
films in these batches were homogeneous without grain boundaries. Magneto-optical 
measurements confirm the thickness dependence of the critical 
current density, as depicted in Figs.~\ref{fig:jc-vs-thickness}(b) and 
\ref{fig:jc(0)-compilation}.
For small film thicknesses, $j$ increases with $d$, 
reaching a plateau for $d \sim 200 $ nm. The current density 
decreases again for the 500 nm--thick film, perhaps due to the 
deteriorating epitaxy as one increases the film thickness too much. 
The increase of the critical current density with the film thickness 
is observed  at all temperatures. The only exception are the data for

\begin{figure}
\centerline{\epsfxsize 8.5cm \epsfbox{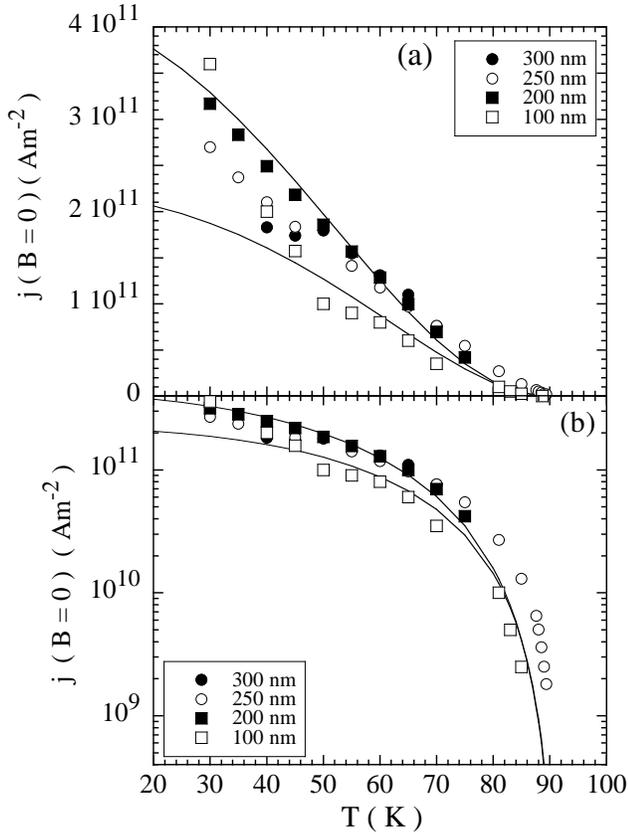}}
\vspace{1mm}
\caption{\label{fig:jc0-vs-T} (a) Temperature dependence of the 
low-field sustainable current density of laser-ablated 
YBa$_{2}$Cu$_{3}$O$_{7-\delta}$ films of 
thicknesses 100 nm $ < d < 300$ nm, deduced from fits to 
Eq.~(\protect\ref{eq:fluxfront}). The temperature dependence can 
be very well fitted  to Eqs.~(\protect\ref{eq:jthinthin}) (lower 
curve) and (\ref{eq:jSV}) (upper curve -- see text). Parameter 
values are $\lambda_{ab}(0) = 120$ nm, $\xi(0) = 2$ nm, the defect 
sizes $D_{i} = 15$ nm and $D_{i}^{z} = 10$ nm,  $B_{c} = 1$ T, 
and a zero-temperature elementary pinning 
force $f_{p,max}^{i}(0) = 8 \times 10^{-11}$ N. (b) the same, on a 
semi-logarithmic scale, in order to bring out the behavior near 
$T_{c}$. }
\end{figure}

\noindent the 150--nm thick films, for which $j$ tends to the values measured in 
thicker films at $T < 0.4 T_{c}$ ( open diamonds in Fig.~\ref{fig:jc(0)-compilation}). 
A similar low--temperature behavior can be noted in Fig.~1 of Ref.~\onlinecite{Griessen94}. 
This may either indicate that a background pinning by other types of defects such 
as dislocations or oxygen vacancies starts to play a role for $T 
\lesssim 35$ K, or that the crossover thickness $d^{*}$ decreases with 
decreasing temperature, a behavior that may be expected from 
Eqs.~(\ref{eq:fp}) and (\ref{eq:d*}).

The temperature dependence of the low--field critical current density 
can be rather well described  using Eq.~(\ref{eq:jthinthin}) and the 
same parameter values as above, as indicated by the lower continuous 
lines in Fig.~\ref{fig:jc0-vs-T} (a,b) (for $d = 100$ nm). Fits using 
Eq.~(\ref{eq:jSV}) for larger film thicknesses
are equally successful [upper lines in Fig.~\ref{fig:jc0-vs-T}(a,b)], 
although $n_{i} = 3 \times 10^{21}$ m$^{-3}$ instead of $n_{i} 
= 1\times 10^{21}$ m$^{-3}$ should be used.

\begin{figure}
\centerline{\epsfxsize 9cm \epsfbox{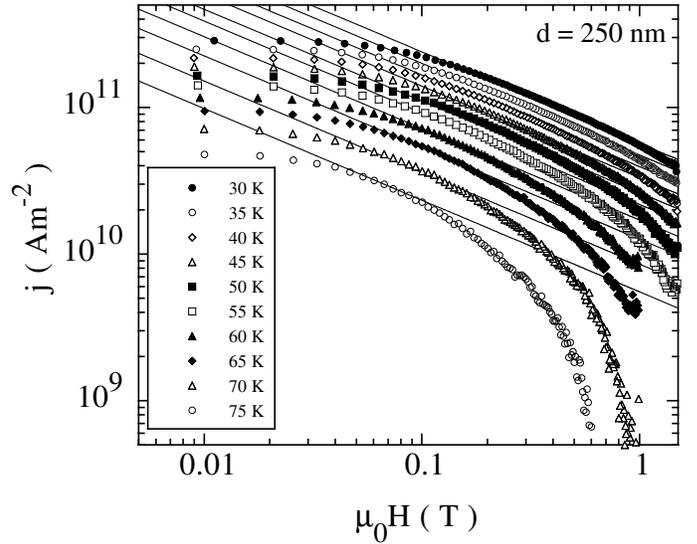}}
\vspace{1mm}
\caption{\label{fig:jc-vs-B} Field dependence of the screening current 
density of a laser-ablated YBa$_{2}$Cu$_{3}$O$_{7-\delta}$ film of 
thickness 250 nm, at different temperatures 55 $ < T < $ 75 K. 
Continuous lines indicate model fits to $j \propto B^{-5/8}$ (Eq~(\protect\ref{eq:jc-strong})); the 
power-law prefactor values are plotted in Fig.~\protect\ref{fig:power-law-prefactor}. }
\end{figure}

\begin{figure}
\centerline{\epsfxsize 8.5cm \epsfbox{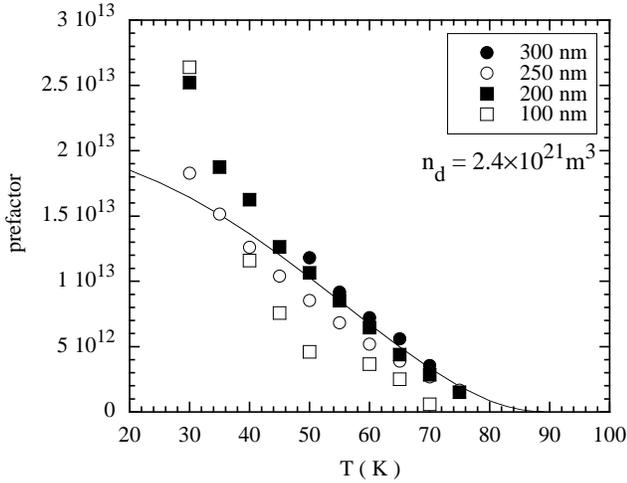}}
\vspace{1mm}
\caption{\label{fig:power-law-prefactor} Prefactors $n_{i}u_{0}^{2} 
f_{p,max}^{i}\Phi_{0}^{-1}B^{5/8} =$ $n_{i} f_{p,max}^{i}\Phi_{0}^{-1} (U_{p}/\varepsilon\varepsilon_{0})^{5/4} 
(\Phi_{0}^{5/8}/\xi^{1/2})$ obtained from the fits in 
Fig.~\protect\ref{fig:jc-vs-B}. the drawn line indicates a fit to 
Eq.~(\protect\ref{eq:jc-strong}) using the same parameter values as 
in Fig.~\protect\ref{fig:jc0-vs-T} ($\lambda_{ab}(0) = 120$ nm, 
$\xi(0) = 2$ nm, $B_{c} = 1$ T, $D_{i} = 15$ nm, $D_{i}^{z} = 10$ 
nm) and $n_{i} = 2.6 \times 10^{21}$ m$^{-3}$. }
\end{figure}

We now turn to the field dependence of the screening current. Typical 
results are depicted in Fig.~\ref{fig:jc-vs-B}, 

\noindent which shows data for the  
250 nm--thick film.  These are in every aspect 
representative of the data for all other films, and very similar to 
those found previously.\cite{Douwes,Gerber91,Dam99} The sustainable 
current density shows a low--field plateau of constant current, followed by a smooth decrease 
of $j$ for fields higher than a given threshold. This 
decrease was recently interpreted in terms of creep of unpinned vortices 
through the forest of vortices presumably trapped on screw dislocation cores, 
a mechanism that ought to lead to a $B^{-1/2}$--dependence of the screening current
density. However, the same field dependence is easily explained in 
terms of strong pinning, see Eqs.~(\ref{eq:jc-strong}) and 
(\ref{eq:jc-cage}). The straight lines in Fig.~\ref{fig:jc-vs-B} show 
fits to the $B^{-5/8}$--dependence predicted by Ovchinnikov and 
Ivlev; the power--law prefactors are given in Fig.~\ref{fig:power-law-prefactor}.
Adapting the same parameter values as for the fits to the zero-field 
 current density, one sees that the temperature dependence of the prefactor 
withstands the comparison with the prediction (\ref{eq:jc-strong})
rather well. Again using the same parameters for the
YBa$_{2}$Cu$_{3}$O$_{7-\delta}$ material and the size of the 
Y$_{2}$O$_{3}$ inclusions, we obtain the drawn line of 
Fig.~\ref{fig:power-law-prefactor} for $n_{i} = 2.6 \times 10^{21}$ 
m$^{-3}$, very close to the previously obtained values. While the 
temperature dependence predicted by Eq.~(\ref{eq:jc-strong}) well 
matches the data at higher temperatures, it fails at temperatures 
below $ \sim 35$ K, again indicating that pinning by other mechanisms may 
become important there.

Note that the field dependence of the sustainable current density 
does not follow a pure power law by any means, but gradually bends 
over from a $B^{-5/8}$--behavior to a $1/B$ dependence. This 
dependence is again in agreement with the strong pinning scenario: 
at high fields the vortex displacements due to pinning become 
comparable to the vortex spacing and the critical current density

\noindent starts do decrease more rapidly, as $j_{c} \propto B^{-1}$ 
(Eq.~(\ref{eq:jc-strong-high-field})). The magnitude of the field 
$B_{a}$ at which this effect is expected is typically of the order 1 T.

\section{Discussion}

The above results show that, even if perfect agreement is not found, 
strong pinning by Y$_{2}$O$_{3}$ inclusions can very well explain the 
large magnitude, as well as the temperature, field, and thickness 
dependence of the critical current density in laser ablated 
YBa$_{2}$Cu$_{3}$O$_{7-\delta}$ films. The primary indication for this 
is the observed dependence on film thickness of the screening current 
density: $j$ first increases as function of film 
thickness before saturating at $d \approx 250$ nm, and eventually 
slowly falling off again. This dependence is very similar to that measured by Jooss 
{\em et al.}.\cite{Jooss96} Other authors also report the increase of $j$ as 
function of film thickness $d$, but with a maximum at $d \approx 100$ 
nm followed by a much more rapid drop of $j$.\cite{Klaassen2002,Liu93,Sheriff97} 
Proposed explanations\cite{Klaassen2002,Jooss96} have been the increase of the effective 
penetration depth from $\lambda_{ab}$ towards $2\lambda_{ab}^{2}/d$ for films 
thinner than $2 \lambda_{ab}$,\cite{Pearl64} a bad film 
morphology and ill-connected islands for $d < 100$ nm,\cite{Liu93} 
and differences in defect structure between the thinner and the thicker 
films.\cite{Sheriff97} We note that the fact that films from different sources display 
maximum $j$ for widely different film thicknesses, 100 nm and 250
nm respectively, exclude that the relative magnitude of $d$ and 
$\lambda_{ab}$ is at the origin of the increase of $j(d)$. However, 
our model explains this difference quite naturally in terms of 
Eq.~(\ref{eq:d*}) and the different density of second-phase 
inclusions in different films.  An effective penetration depth
$2 \lambda_{ab}^{2} /d$ would modify the current density obtained from
pinning by thickness variations, but  \em not \rm the pinning force in the 
case of core pinning, for instance, by extended point defects or linear defects.
The differences in optimal thickness also seem to exclude bad film 
morphology as an explanation: while such an effect may be likely for 
very thin films of thickness much less than 100 nm, it cannot explain 
an increase of $j(d)$ up to $d = 200$ nm. Finally, the role of 
deep trenches that modify the total current circulating in thin 
films\cite{Sheriff97} is excluded by the magneto-optical observations, 
which show homogeneous, featureless flux penetration. Thus, we 
conclude that the observed thickness dependence is in 
agreement with the scenario of strong pinning by extended point 
defects {\em only}. 

Further evidence is the the lack of a cusp--like angular dependence of 
the critical current density for field aligned close to the film 
normal, which argues against a 
predominant role of extended line-like defects such as screw 
dislocation cores. The large difference between the  
algebraic temperature dependence of $j_{c}$ with the exponential 
dependence observed in clean YBa$_{2}$Cu$_{3}$O$_{7-\delta}$ single 
crystals renders the relevance of weak collective pinning by oxygen 
vacancies unlikely. These observations gain in importance if we note that the field and
temperature dependences reported here are observed in YBa$_{2}$Cu$_{3}$O$_{7-\delta}$ films 
quite irrespective of the deposition method and conditions. While 
generally neglected, small Y$_{2}$O$_{3}$ inclusions are ubiquitous 
in YBa$_{2}$Cu$_{3}$O$_{7-\delta}$ films deposited by different 
methods and may provide an explanation for both the magnitude and 
behavior of $j_{c}$ and the generality of this behavior. In our case, 
the defect density needed to explain the experimentally observed 
$j_{c}$, $1\times 10^{21} < n_{i} < 3 \times 10^{21}$ m$^{-3}$ means 
that the precipitates occupy between $0.2\%$ and $0.6\%$ of the sample 
volume. The $5 \%$ of secondary phase material found in the 
X-ray analysis would be present as larger CuO$_{2}$ outgrowths which do not 
pin vortices efficiently.

Strong pinning by larger point--like inclusions dominates over 
collective pinning by oxygen vacancies because thermal fluctuations in 
high temperature superconductors efficiently smear out the pinning 
potential of the vacancies.\cite{Feigelman90} Even at zero 
temperature it is energetically more favorable to pin a vortex on 
rather large inclusions than to have it lower its energy by wandering 
through the weak collective pinning landscape. Restricting ourselves 
to the single vortex limit, pinning on a large inclusion yields an energy gain 
$\frac{1}{4}\varepsilon_{0}D_{i}^{z}$; this should be compared to the 
potential energy gain due to pinning by oxygen vacancies in the 
volume $u_{0}^{2}\overline{\cal L}$. Strong pinning by Y$_{2}$O$_{3}$ 
inclusions is more favorable if

\begin{equation}
	\frac{1}{4}\varepsilon_{0}D_{i}^{z} > {\cal 
	U}_{p}\left(\frac{\overline{\cal L}}{L_{c}}\right)^{2\zeta-1}
	\end{equation}

\noindent where ${\cal U}_{p}$ is the pinning energy gained in a 
correlation volume due to weak collective pinning by the oxygen 
vacancies and $\zeta \approx 0.63$ is the wandering exponent.\cite{Blatter94} 
Substituting  $\overline{\cal L}$ from Eq.~(\ref{eq:L-SV}), assuming ${\cal U}_{p} = 
\varepsilon_{0}(\frac{1}{16}\varepsilon^{2}n_{v}D_{v}^{4}\xi^{2})^{1/3}$ 
for $\delta \kappa$--pinning (pinning by mean free path fluctuations), and the 
oxygen ion radius $D_{v} \approx 3\times 10^{-10}$ m, we find that for 
physically allowable oxygen vacancy densities $n_{v}$ strong pinning 
by Y$_{2}$O$_{3}$ inclusions is always more favorable.

In the above, we have disregarded the effects of flux creep. While this 
can more or less be justified at low temperatures where creep rates 
are small, at higher temperatures flux creep certainly determines the 
observed sustainable current density. Experiments by Klaassen {\em et 
al.} \cite{Klaassen2000}
show that in the low--field regime of constant current density the 
activation energy for flux creep $E_{a}$ is of the order of 600 K; the 
activation barrier rapidly decreases and becomes constant at fields 
where $j$ decreases as function of $B$. This low value of $E_{a}$, 
much lower than the value predicted by nucleation-type creep models 
for depinning from linear extended defects,\cite{Nelson92,Nelson93}
could be explained by the fact that vortices are held not by extended 
defects but by second-phase inclusions of much shorter longitudinal 
dimensions. The rapid 
decrease of the activation energy with magnetic field would be due to 
the decrease in the average number of defects, by which any given 
vortex line is effectively pinned ({\em i.e.} $\overline{\cal L}$ 
becomes of the order of the film thickness). The high--field regime 
of constant activation barrier can be explained if one assumes that 
there is less than one effective pin per vortex line; Our experimental 
result $n_{i} \approx 1 \times 10^{21}$ m$^{-3}$ implies a distance 
between defects of the order 100 nm, comparable to both the smallest film 
thicknesses and the vortex spacing at $B = 0.2$ T. So, indeed, in the field regime 
of decreasing $j(B)$, there is less than one effective defect per 
vortex, and the flux creep activation energy should be given by the energy 
barrier needed to break free a line
from one defect. The power-law field dependence implies that the 
fraction of vortices pinned in this way suffices to keep the whole 
vortex lattice at rest; at the threshold (critical) current density, 
the entire vortex lattice depins until a vortex line is trapped by 
the next defect. Only at higher fields, where the experimental 
$j(B)$--data significantly deviate from a power--law, do we
expect creep effects, vortex lattice shear, and plastic deformations 
of the vortex lattice to become important.

\section{Conclusions}

We have measured the dependence of the sustainable current density in pulsed-laser deposited 
YBa$_{2}$Cu$_{3}$O$_{7-\delta}$ thin films on film thickness, field orientation 
and magnitude, and temperature. While the field and 
temperature dependence of the current density show the  
qualitative behavior often reported in literature for laser 
ablated, sputtered, and MO-CVD thin films, the 
film thickness dependence is consistent only with strong pinning by 
sparse large second phase inclusions (such as Y$_{2}$O$_{3}$). A model 
description of such pinning in different limits of flux line density 
and film thickness is consistent with the experimentally found 
behavior, if one assumes the inclusions to have the typical size $15 \times 15 
\times 10$ nm$^{3}$ obtained from previous TEM studies,
\cite{Selinder92,Verbist96}, with a defect density $n_{i} \sim 1 - 3
\times 10^{21}$ m$^{-3}$. Such a defect density suggests in turn
that at high fields, only a fraction of the vortices are pinned; these 
suffice to hold the entire lattice at rest until the critical current 
density is reached.

\nonumber\section{Acknowledgements} This work was partially supported by Polish Government
grant No.'' PBZ-KBN/013/T08/19. Support from French - Polish Programme ``Polonium'' and ESF
``VORTEX programme is also gratefully acknowledged.


\end{multicols}
\end{document}